\newcommand{\<}{\langle}
\renewcommand{\>}{\rangle}                    
\newcommand{\beq}{\begin{equation}}
\newcommand{\eeq}{\end{equation}}
\newcommand{\beqn}{\begin{eqnarray}}
\newcommand{\eeqn}{\end{eqnarray}}
\newcommand{\nn}{\nonumber}

\documentstyle[12pt]{article}
\begin{document}
\begin{flushright} 
BUHEP-01-7\\ 
CERN-TH/2001-130\\ 
ROME1-1315/2001\\ 
ROM2F/2001/14\\
\end{flushright} 
\centerline{\LARGE{\bf The $U_A(1)$ Problem on the Lattice}}
\vskip 0.3cm
\centerline{\LARGE{\bf with Ginsparg-Wilson Fermions}}
\vskip 1.0cm 
\centerline{ {{\large{L. Giusti}}\,$^{{a)}}$ 
{\large{G.C. Rossi}}\,$^{{b)}}$\footnote{On leave 
of absence from Dipartimento di Fisica, Universit\`a di Roma 
``{\it Tor Vergata}'', Via della Ricerca Scientifica, 
I-00133 Roma, Italy.} {\large{M. Testa}}\,$^{{c)}}$
{\large{G. Veneziano}}\,$^{{b)d)}}$}}
\vskip 0.5cm 
\centerline{$^{a)}$\small{Physics Department, Boston University,}}
\centerline{\small{590 Commonwealth Avenue, Boston, MA 02215, USA}}
\smallskip 
\centerline{$^{b)}$\small{Theory Division, CERN, 1211 Geneva 23, 
Switzerland}}
\smallskip 
\centerline{$^{c)}$\small{Dipartimento di Fisica, Universit\`a di Roma
``{\it La Sapienza}",}}
\centerline{\small{INFN, Sezione di Roma ``{\it La Sapienza}''}}
\centerline{\small{P.le A. Moro 2, I-00185 Roma, Italy.}}
\smallskip 
\centerline{$^{d)}${\small{Laboratoire de Physique Th\'eorique, 
Universit\'e Paris Sud, 91405 Orsay, France}}}
\vskip .5cm 
\centerline{\bf ABSTRACT} 
\begin{quote} 
{\small{We show how it is possible to give a precise and unambiguous 
implementation of the Witten--Veneziano formula for the $\eta'$ mass
on the lattice, which looks like the formal continuum one, if the expression 
of the topological charge density operator, suggested by fermions obeying
the Ginsparg--Wilson relation, is employed. By using recent numerical results 
from simulations with overlap fermions in 2 (abelian Schwinger model) and 4 
(QCD) dimensions, one obtains values for the mass of the lightest 
pseudo-scalar flavour-singlet state that agree within errors with theoretical 
expectations and experimental data, respectively}}.
\end{quote} 
\vfill 
\newpage
\section{Introduction}
\label{sec:INTRO}

Lacking any reliable analytical way to compute bound-states masses in a 
strongly interacting theory, a crucial confirmation of QCD as the theory of 
hadron dynamics has come in recent years from our ability to compute from 
first principles the masses of the lightest hadrons in Lattice QCD
(LQCD)~\cite{W}. This has been done with remarkable success for a
number of mesonic and baryonic states, both in the quenched approximation 
(no fermion determinant) and, more recently, also in the full (unquenched) 
theory~\cite{AOK}.

A somewhat difficult problem is, computationally, the inclusion
of diagrams violating the so-called OZI-rule~\cite{OZI}, i.e.~of
diagrams involving purely gluonic intermediate states: these require
the calculation of the trace of the inverse Dirac operator, and are
affected by large statistical fluctuations. The very existence of
such diagrams is a key property of QCD, making it differ in an essential 
way from any naive quark model. Since OZI-violating diagrams are the ones 
that distinguish flavour-singlet from flavour-non-singlet states, their 
inclusion in LQCD is crucial and should allow for the theoretical 
computation of subtle phenomena, such as $\rho-\omega-\phi$ and 
$\pi-\eta-\eta'$ splitting and mixing. In fact, one of the most striking 
experimental facts in hadronic physics is the contrast between the vector 
(as well as the tensor) mesons, that appear to satisfy the OZI-rule to 
high accuracy, and the pseudo-scalar mesons, whose (mass)$^2$ eigenstates are 
very far from being ideally mixed (the OZI-prediction), and whose eigenvalues
exhibit a peculiar hierarchical structure, $m_{\pi}^2 \ll m_{\eta}^2 
\ll m_{\eta'}^2$.

The technical problems mentioned above  have so far prevented in LQCD 
a fully-fledged evaluation of the $\eta'$ mass at the same level of 
accuracy as that reached for the other light hadrons, although noticeable 
progress has been recently made in this direction~\cite{AAK,SSUE}.

In quenched LQCD the first evaluation of the $\eta'$ 
mass~\footnote{Throughout this paper we will indicate by  
$\eta'$ the flavour-singlet pseudo-scalar meson, ignoring vector flavour 
symmetry-breaking effects. We will work in Euclidean metric with hermitian
$\gamma$ matrices satisfying $\{\gamma_\mu,\gamma_\nu\}=2\delta_{\mu\nu}$ and
$\mbox{tr}(\gamma_5\gamma_1\gamma_2\gamma_3\gamma_4)= 4\epsilon_{1234}=4$. 
Left-handed and right-handed fermions are defined through the equations
$\gamma_5 \psi_L=\psi_L$ and $\gamma_5 \psi_R=-\psi_R$, respectively.
Colour generators, $T^a$ ($a=1,...,N_c^2-1$), in the fundamental 
representation are normalized according to ${\rm{tr}}(T^aT^b)=
\delta^{ab}/2$.} dates back to the work of ref.~\cite{PAR}, where 
the OZI-violating (ZV) and the OZI-conserving (ZC) contributions to the 
2-point meson--meson Green function
\beq
\Gamma_{q}(t) = \int d{\vec {x}}\,\<P^{0}({\vec {x}},t)
P^{0}(0)\>\Big{|}_{\rm{quenched}} \label{P5P5} \eeq
were computed. In eq.~(\ref{P5P5})
\beq
P^{0} = \sum_{r=1}^{N_f} \bar\psi^r\gamma_5\psi_r
\label{PO}\eeq
is the pseudo-scalar singlet quark density, with $N_f$ the number of light
quarks. The (quenched value of the) $\eta'$ mass is identified as the amount 
by which the singlet pseudo-scalar $\bar q q$-boson pole gets shifted owing 
to the iteration of Nambu--Goldstone (NG) particle exchanges, the sum of 
which builds up the full (unquenched) expression of $\Gamma$. This leads 
to the formula
\beq
m_{\eta'}^2=2 m_{\pi}\lim_{t\rightarrow\infty}\frac{\Gamma^{\rm{ZV}}_{q}(t)}
{|t|\Gamma^{\rm{ZC}}_{q}(t)}\, .
\label{ETAPAR}\eeq
Numerical data were encouraging, but far from convincing. Although
considerable progress has been achieved along these lines in more recent 
simulations~\cite{KUR,KIL}, in our opinion one cannot regard the
present numerical results as really conclusive.

From a more theoretical point of view, leaving aside the lattice for a
moment, the striking difference between vector and pseudo-scalar
mesons appeared as a serious problem within the generally
accepted framework according to which the light pseudo-scalar
mesons are the (quasi) NG bosons of a spontaneously (and also explicitly) 
broken chiral symmetry. As first noted by Glashow~\cite{GLA}, a  naive use of
spontaneously-broken chiral symmetry calls for nine - rather than
eight - light NG bosons ($N_f=3$). Thus one would naively expect that chiral 
symmetry should ``protect" the OZI-rule in the pseudo-scalar channel, 
while the opposite is experimentally true. 

Awareness of the problem was reinforced  by a very influential paper
by Weinberg~\cite{WEIN} in which, under
what he calls ``plausible assumptions'', the bound
\beq
m_{{\rm{octet}}, I=0} < \sqrt{3} m_{\pi}
\label{WBOUND} \eeq
was derived in the absence of $U_A(1)$ anomaly contributions.

Since the very beginning, the famous ABJ anomaly~\cite{ABJ} was suspected to 
have much to do with the resolution of the $U_A(1)$ puzzle, as the strong
(QCD) axial anomaly appears only in the flavour-singlet channel. But the 
fact that the anomaly itself is a total divergence (albeit of a 
non-gauge-invariant current) made it look irrelevant for the solution of the 
problem. Kogut and Susskind~\cite{KSS} were the first to notice that the ABJ 
anomaly could be made effective again, if one assumed the existence of a 
``ghost" particle, a massless state coupled to the non-gauge-invariant 
current, but decoupled from any gauge-invariant operator.

Things took a decisive turn when, in 1976, 't Hooft~\cite{THOO} pointed 
out that the resolution of the $U_A(1)$ problem had to be related to the 
existence of topologically non-trivial gauge field configurations
(instantons) in Euclidean QCD. Soon after, Crewther~\cite{CRE} emphasized  
that 't Hooft's crucial observation could only be considered a solution 
of the problem, if it could pass a number of theoretical tests, in particular 
that of satisfying both the conventional (i.e.~non-anomalous) and anomalous 
Ward--Takahashi Identities (WTI's). Several puzzles, related to the 
dependence on the number and masses of quark flavours appeared, in fact, to 
be still present~\cite{CRE}. 

Fortunately, these questions can be addressed and solved by exploiting the 
simplifications that occur in certain limits of QCD. One such limit, employed 
in ref.~\cite{WIT}, is the 't~Hooft limit ($N_c\rightarrow\infty$, with 
$g^2 N_c$ and $N_f$ held fixed~\cite{NINF}). The other possibility, proposed 
in ref.~\cite{VEN}, proceeds by assuming that anomalous flavour-singlet axial 
WTIs retain their validity order by order in an expansion in 
$u\equiv N_f/N_c$ around $u=0$~\footnote{This expansion is not to be 
confused with the topological expansion ($N_c\rightarrow\infty$ at fixed 
$g^2 N_c$ and $u$), introduced in ref.~\cite{VENTOP}.}. In either cases 
one can derive the leading-order Witten--Veneziano (WV) relation
\beq
m_{\eta '}^2=\frac{2N_f}{F_\pi^2}A\, ,
\label{ETAMASS}\eeq
where $F_\pi\simeq 94\,{\rm{MeV}}$ is the pion decay constant
and $A$ is the ``topological susceptibility". $A$ is formally  defined 
by the equation
\beq
A=\int d^4x\, \<Q(x) Q(0) \>\Big{|}_{\rm{YM}}\, ,
\label{A}\eeq
where $Q(x)$ is the topological charge density
\beq
Q(x)=\frac{g^2}{32\pi^2}\epsilon_{\mu\nu\rho\sigma}
{\rm{tr}}[F_{\mu\nu}F_{\rho\sigma}(x)] 
\label{Q}\eeq
and the trace is over colour indices. The notation $\<...\>|_{\rm{YM}}$
in eq.~(\ref{A}) means that the $QQ$-correlation function is to be
computed in the pure Yang--Mills (YM) theory, i.e.~in the absence of quarks.

We recall that, within the approach of ref.~\cite{WIT}, eq.~(\ref{ETAMASS}) is
expected to be valid for large $N_c$, while, relying on the derivation  
given in ref.~\cite{VEN}, one would conclude that the $\eta'$ mass formula is  
valid for any value of $N_c$ at  leading order in $u$. We also notice that the
limit $u\rightarrow 0$ at fixed $g^2N_c$ encompasses the 't~Hooft limit
($N_c\rightarrow\infty$ at fixed $N_f$) and that the equivalence of  
eqs.~(\ref{ETAPAR}) and~(\ref{ETAMASS}) essentially relies on the validity 
of anomalous flavour-singlet WTIs order by order in $u$~\cite{VEN}.

Since the pioneering works of refs.~\cite{THOO},~\cite{WIT} and~\cite{VEN},
a large number of papers (see, for instance,~\cite{DIV}--\cite{TEP}
and references therein) have appeared, aimed at checking the validity of
eq.~(\ref{ETAMASS}) by comparing the number obtained by inserting in it the 
value of $A$ extracted from pure gauge lattice simulations with the 
experimental magnitude of the $\eta'$ mass~\footnote{The 
question of how eq.~(\ref{ETAMASS}) should be modified beyond the chiral 
limit was addressed in refs.~\cite{VEN,NIC}.}.

The formal eqs.~(\ref{ETAMASS}) and~(\ref{A}), when translated in any
regularized version of QCD, such as LQCD to which we will refer in the
following, become more complicated and quite a number of subtleties have to
be dealt with in order to be able to determine the correct field theoretical 
expression of $A$, which should be inserted in the $\eta'$ mass formula.

Two main problems need to be solved to make eqs.~(\ref{ETAMASS})  
and~(\ref{A})
rigorous and of practical use. One has to
i) find a properly normalized lattice definition of the topological
charge density $Q_L$;
ii) subtract from the $Q_L(x)Q_L(0)$ operator product appropriate
contact terms, as required to make it an integrable (operator-valued)
distribution.

Three types of prescriptions have been proposed upon addressing the first
problem. They can be succinctly described as follows.
\begin{itemize}
\item Take any naive lattice discretization of
$F_{\mu\nu}$~\cite{DIV}~\cite{BAA}~\cite{CHR} and construct $Q_L$ through 
(the lattice version of) eq.~(\ref{Q}) by properly normalizing it~\cite{DIG}.
\item Define $Q_L$ by making reference to the anomalous flavour-singlet axial
WTIs~\cite{YOS}~\cite{UNG}~\cite{OVER}.
\item Exploit the topological nature of $Q$ to give a definition of it
appropriate to a discrete manifold~\cite{LU1}.
\end{itemize}

The second problem is actually more subtle than stated above. In fact,
it is not sufficient to subtract (possible) {\it divergent} contact terms  
from the r.h.s. of eq.~(\ref{A}), by taking properly into account the mixing 
of $Q(x)Q(0)$ with lower dimensional operators. It is also necessary to fix 
{\it finite} contact terms of the form, say, $c\,\delta(x)$, since, as 
also stressed in ref.~\cite{CRE}, they would contribute a constant to 
the $\eta'$ mass when inserted in eq.~(\ref{A}).

The question of the subtraction of lower dimensional operators was addressed
in an incomplete way and in perturbation theory in ref.~\cite{DIV}, and in a
more refined, non-perturbative, way in ref.~\cite{DIG}. The issue was taken 
up again more recently within the overlap 
formulation of LQCD~\cite{NEUB,OVER} (see ref.~\cite{NI} for a recent review).

The purpose of this paper is to give a rigorous derivation of a formula
for the $\eta'$ mass, which could be unambiguously used in numerical
simulations. The validity of such a formula is a crucial test 
of QCD, as it directly relates the non-vanishing of the 
$\eta'$ mass in the chiral limit to the explicit breaking of the $U_A(1)$ 
symmetry induced at the quantum level by the gluon anomaly.

Our strategy is to use (lattice) regularized anomalous flavour-singlet axial 
WTIs of full QCD in order to properly construct the $\<QQ\>$ correlation 
function, as a well defined (integrable) distribution. The remarkable result 
of our analysis is that no subtraction is needed to this end in 
the chiral limit, if the definition of lattice topological charge 
density suggested by fermions obeying the Ginsparg--Wilson (GW) 
relation~\cite{GW} (e.g.~overlap fermions~\cite{NEUB,OVER}) is employed. 
In this case a formula for the $\eta'$ mass, which looks like the naive 
continuum one (i.e.~which closely parallels the structure of 
eqs.~(\ref{ETAMASS}) and~(\ref{A})), can be obtained.

The case of Wilson fermions is somewhat more involved and will be 
discussed in a forthcoming publication~\cite{GRTV}.

The paper is organized as follows. In sect.~\ref{sec:CONT} we recall the 
derivation of the $\eta'$ mass formula in the continuum. In 
sect.~\ref{sec:GWF} we show how one can give a precise meaning to this 
relation in the QCD regularization offered by GW fermions. 
In sect.~\ref{sec:NUM} we discuss 
the numerical evidence in favour of the WV formula in 2 and 4 dimensions, 
which emerges from the existing overlap fermion simulation data. 
Conclusions can be found in sect.~\ref{sec:CONCL}. In an Appendix we derive 
several relations between unquenched and quenched QCD quantities, involving 
moments of the topological susceptibility, including the WV formula. 

\section{The $\eta'$ mass formula in the continuum}
\label{sec:CONT}

In this section we review the formal derivation of the WV formula in 
the continuum. We start from the anomalous flavour-singlet WTI
\beq
\partial_\mu\<A_\mu^0(x) Q(0)\>=
2 m_q\<P^0(x) Q(0)\>+ 2 N_f\<Q(x) Q(0)\> \, ,
\label{WTIC} 
\eeq
where
\beq
A_\mu^0 = \sum_{r=1}^{N_f}
\bar\psi^r\gamma_\mu\gamma_5\psi_r
\label{AXCUR}\eeq
is the axial current, $P^0$ was defined in eq.~(\ref{PO}), $Q$ is the 
topological charge density and $m_q$ is the quark mass. It is understood
that all operators in eq.~(\ref{WTIC}) are finite (i.e.~have finite
insertions with any string of renormalized fundamental fields), so that 
every term in eq.~(\ref{AXCUR}) is finite as soon as $x\neq 0$.

The Fourier transform of eq.~(\ref{WTIC}) in the chiral limit reads
\begin{eqnarray}
& &  i p_\mu \int d^4 x \, \mbox{e}^{-ipx} 
 \<A^0_\mu(x) Q(0)\> +{\rm{CT}}(p) =\label{GQQNC} \\
& & = 2 N_f\int d^4 x \, \mbox{e}^{-ipx}\<Q(x) Q(0)\>+{\rm{CT}}(p)\equiv
2 N_f\chi_t(p) \, .\nonumber
\end{eqnarray}
A contact term, ${\rm{CT}}(p)$, has been added to both sides of 
eq.~(\ref{GQQNC}) to make them separately finite. ${\rm{CT}}(p)$ is a 
polynomial of degree 4 in $p$, which vanishes at $p=0$, because the Fourier
transform of the l.h.s. of eq.~(\ref{WTIC}) is certainly finite (actually 
zero) at $p=0$ in the full theory. As will become clear in the following, 
${\rm{CT}}(p)$ plays no r\^ole in the main line of arguments we present in 
this paper, since finally we will be only interested in the value of $\chi_t$ 
at $p=0$~\footnote{The presence of ${\rm{CT}}(p)$ will, instead, affect the 
determination of quantities like $\chi'_t(0) = d\chi_t(p^2)/dp^2|_{p=0}$, 
unless ${\rm{CT}}(p)\propto p^4$~\cite{SHO}. $\chi'_t(0)$ is relevant to 
the discussion of the so-called proton spin crisis problem~\cite{PSCP}. 
We will elaborate further on this issue in the Appendix.}.

The absence of zero-mass particles in the pseudo-scalar singlet channel
implies the ``sum rule''
\beq
\chi_t(0) = 0\, .
\label{SUM}\eeq
This relation is the famous statement that the topological susceptibility
in full QCD vanishes in the chiral limit. 

One can derive from it a formula for the $\eta'$ mass whenever the 
topological susceptibility in the absence of quarks is non-zero. In this 
situation, in fact, the vanishing of $\chi_{t}(p)$ at $p=0$ in full QCD 
to any order in $u$, can possibly take place only if there exists a 
$\bar q q$-meson contributing to $\chi_{t}(p)$, whose mass goes to zero 
(linearly) as $u\rightarrow 0$~\cite{WIT}~\cite{VEN}. Under these 
circumstances the limit $p\rightarrow 0$ and the expansion of 
$\chi_{t}(p)$ around $u=0$ do not commute. In the Appendix we give a 
derivation of the WV formula along these lines and prove a number of other 
interesting related equations. The formulae that are obtained are derived
under the assumption that at $p\neq 0$ all the relevant quantities and 
correlators are smooth functions of $u$ and, in particular, that taking 
the limit $u\rightarrow 0$ is equivalent to neglecting the fermion determinant.
In the following, for short, we will refer to this assumption as the 
``smooth-quenching hypothesis''.

To leading order in $u$ one then obtains for the $\eta'$ mass either
the formula
\beq  m^2_{{\eta}'}=
\frac{1}{F_\pi^2} \int d^4 x \,
\partial^\mu\<A_\mu^0(x)Q(0)\>\Big{|}_{\rm{quenched}} \, ,
\label{ETAUSE} \eeq
or, equivalently
\beq
m^2_{{\eta}'}=
\frac{2N_f}{F_\pi^2} \int d^4 x \, 
\<Q(x) Q(0)\>\Big{|}_{\rm{YM}} \, . 
\label{ETAFIN} \eeq
We stress that, under the above assumptions, the r.h.s. of (\ref{ETAFIN}) 
(as well as that of eq.~(\ref{ETAUSE})) is ultraviolet-finite, 
because ${\rm{CT}}(p)$ is a polynomial in $p$ that vanishes at $p=0$
for any value of $u$.

In the next section we will show how it is possible to give a precise meaning 
to the above equations, starting with the chirally invariant regularization 
of QCD offered by fermions obeying the GW relation. In this framework 
we will be able to derive an unambiguous formula for the $\eta'$ mass, 
immediately usable in actual Monte Carlo simulations.

We conclude by remarking that eq.~(\ref{ETAUSE}) cannot be employed for direct 
numerical evaluations, because its r.h.s. is actually identically zero 
at finite volume, as there are no massless particles in a finite box. However, 
it can be put in a form that is amenable to numerical simulations, if one 
goes one step further and rewrites it in the form~\footnote{See the Appendix 
for the relevant normalizations and notations.}
\beq
m^2_{{\eta}'}=
\frac{\sqrt{2N_f}}{F_\pi}\<0|Q|\eta'\>\Big{|}_{\rm{quenched}} \, .
\label{ETAQ} \eeq
This equation follows from the fact that in the quenched limit the integral 
in the r.h.s. of eq.~(\ref{ETAUSE}) is exactly given at the chiral point by 
the contribution of the pole of the singlet pseudo-scalar particle, as the 
mass of the latter vanishes in that limit. No explicit reference to the size 
of the volume in which the system is enclosed appears anymore in 
eq.~(\ref{ETAQ}). We will make use of these considerations in 
sect.~\ref{subsec:ETA}.

\section{Ginsparg-Wilson fermions}
\label{sec:GWF}

Regularizing the fermionic part of the QCD action using GW
fermions~\cite{GW,NEUB,OVER} offers the great advantage over the standard
Wilson discretization~\cite{W} that global chiral transformations can be 
defined, which are an exact symmetry of the massless theory, as in the formal  
continuum limit. This is a consequence of the relation~\cite{GW}~\footnote{We 
set the lattice spacing, $a$, equal to 1. It is, however, understood that 
in all lattice formulae the continuum limit should be finally taken.}
\begin{equation}
\gamma_5D+D\gamma_5=D\gamma_5D\, ,
\label{GWREL}
\end{equation}
obeyed by the GW Dirac operator, $D$. In this regularization the $U_A(1)$ 
anomaly is recovered \`a la Fujikawa~\cite{F}, as a consequence of the 
non-trivial Jacobian that accompanies the change of fermionic integration 
variables induced by a $U_A(1)$ lattice transformation. In fact, it can be 
shown~\cite{HLNL} that under the infinitesimal global transformations
\beqn
& & \psi=\psi'+\epsilon\gamma_5 (1 - D)\psi'
\nn \\
& & \bar\psi=\bar\psi'+
\epsilon\bar\psi' \gamma_5
\label{UACOV} \eeqn
the massless action, thanks to~(\ref{GWREL}), remains invariant, while the 
functional integration measure gets multiplied by the factor
\begin{equation}
J=\exp\{\epsilon \int d^4x\, {\rm{Tr}}[\gamma_5 D(x,x)]\}\, ,
\label{JAC}
\end{equation}
where Tr is over colour and spin indices. In view of these results
it is concluded that~\footnote{Actually the GW relation~(\ref{GWREL}) 
alone may not be enough to ensure the validity of eqs.~(\ref{INDEX}) 
and~(\ref{CHARDEN}) below~\cite{CHI}. These equations are, however, valid 
if overlap fermions are used~\cite{ADA} and, more generally, if the 
admissibility condition of ref.~\cite{HJL} is fulfilled.} 
\begin{itemize}
\item
the index theorem for the GW Dirac operator reads (for each flavour)
\begin{equation}
n_R-n_L={\rm{index}}(D)=\frac{1}{2}\int d^4x\, {\rm{Tr}}[\gamma_5 D(x,x)]\, ;
\label{INDEX}
\end{equation}
\item
for sufficiently smooth gauge configurations~\cite{HJL} one has
\begin{equation}
\frac{1}{2} {\rm{Tr}}[\gamma_5 D(x,x)]=
\frac{g^2}{32\pi^2}\epsilon_{\mu\nu\rho\sigma}
{\rm{tr}}[F_{\mu\nu}F_{\rho\sigma}(x)] \, ;
\label{CHARDEN}
\end{equation}
\item
the anomalous flavour-singlet WTIs in the presence of $N_f$ massless
fermions take the form
\begin{equation}
0=\frac{2N_f}{2}\int d^4x\,\<{\rm{Tr}}[\gamma_5 D(x,x)] \hat{O}\>+
\<\delta_A \hat{O}\>\, ,
\label{AWTI}
\end{equation}
where $\hat{O}$ is any finite (multi)local operator. In eq.~(\ref{AWTI}) 
we have symbolically indicated by $\epsilon\delta_A \hat{O}$ the infinitesimal 
variation of $\hat{O}$ under the transformations~(\ref{UACOV}).
Equation (\ref{AWTI}) assumes the absence of a $U_A(1)$ NG boson.
\end{itemize}

Starting with the local version of the transformations~(\ref{UACOV}), 
one gets in the chiral limit the local WTIs
\begin{equation}
\nabla_\mu\<A^0_\mu (x) \hat{O}\>=\frac{2N_f}{2}
\<{\rm{Tr}}[\gamma_5 D(x,x)] \hat{O}\>+\<\delta^x_A \hat{O}\> \, ,
\label{AWTIL}
\end{equation}
where $\delta^x_A \hat{O}$ is the local variation of $\hat{O}$; 
$\nabla_\mu A^0_\mu (x)$ is the divergence of the singlet axial current,
a quantity that formally vanishes when integrated over all space-time.

Neither ${\rm{Tr}}[\gamma_5 D(x,x)]$ nor $A^0_\mu (x)$ are finite operators, 
but finite linear combinations are easily constructed~\cite{BARD,TES}. In 
fact, since the second term in the r.h.s. of eq.~(\ref{AWTI}) is obviously 
finite, if $\hat{O}\rightarrow\hat{\Lambda}_n\equiv \psi_R(x_1)...
\bar{\psi}_R(x_n)$, it follows that the integrated operator 
$\int d^4x\,{\rm{Tr}}[\gamma_5 D(x,x)]$ 
is also finite, as it has finite insertions with any string of renormalized 
fundamental fields. This statement in turn implies that 
${\rm{Tr}}[\gamma_5 D(x,x)]$ can only mix with 
operators of dimension $\leq 4$ and vanishing integral, hence only with 
$\nabla_\mu A^0_\mu(x)$. Calling $Z$ the relevant mixing coefficient, 
one can define the finite operators $\hat Q$ and $\hat A^0_\mu$ by writing
\beq
\hat{Q}(x)=\frac{1}{2}{\rm{Tr}}[\gamma_5 D(x,x)] - 
\frac{Z}{2N_f}\nabla_\mu A_\mu^0 (x)
\label{FINQ1}\eeq
\beq
\hat{A}_\mu^0 (x)=(1-Z) A_\mu^0 (x)\, .
\label{FINA} \eeq
$Z$ is logarithmically divergent to lowest order in perturbation theory and 
vanishes as $u\rightarrow 0$. Its perturbative expansion starts at order $g^4$.

With the above definitions the renormalized singlet axial WTI becomes
\begin{equation}
\nabla_\mu\<\hat A^0_\mu (x) \hat{O}\>=
2N_f\<\hat Q(x) \hat{O}\>+\<\delta^x_A \hat{O}\> \, .
\label{WTIREN}
\end{equation}
We expressly note that 1) as the comparison of the WTIs~(\ref{AWTIL}) 
and~(\ref{WTIREN}) shows, the operators $\hat Q$ and $\hat A^0_\mu$ are 
already correctly normalized, so no further (finite) multiplicative 
renormalization is required; 2) the line of arguments given above also 
proves that there is no mixing of ${\rm{Tr}}[\gamma_5 D(x,x)]$ with operators 
of dimension smaller than 4 (like the pseudo-scalar quark density), which 
would bring in dangerous power divergent mixing coefficients. This fact is 
related to the absence of power-like divergent, $1/a$, subtraction in the 
definition of renormalized quark mass. All these are very distinctive 
features of GW fermions which are at variance with what happens in the case 
of Wilson fermions~\footnote{We are indebted to M. L\"uscher for pointing 
out to us the possible existence of mixing between $Q$ and $P^0$ in the case 
of lattice Wilson fermions.}.

\subsection{The $\eta'$ mass formula on the lattice}
\label{subsec:ETA}

A formula for the $\eta'$ mass can be easily obtained from the 
WTI~(\ref{AWTIL}), assuming that in the limit $u\rightarrow 0$ the $\eta'$ 
mass is O($u$). A more sophisticated derivation can be found in the Appendix.

One starts by defining the lattice Green function
\begin{equation}
\chi_{tL}(p)=\frac{1}{2N_f}\int d^4x\, {\rm{e}}^{-ipx} 
\nabla_\mu\<\hat A^0_\mu (x) \hat Q(0)\>\, ,
\label{AWTI2}
\end{equation}
with $\hat{Q}$ given by eq.~(\ref{FINQ1}). The same observation, that was 
made in sect.~\ref{sec:CONT} after eq.~(\ref{GQQNC}) about the presence of 
counter-terms necessary to make the correlator~(\ref{AWTI2}) finite at 
$p\neq 0$, applies also here. However, as this counter-term does not play 
any r\^ole in the determination of the $\eta'$ mass, for brevity we will not 
indicate it explicitly in the formulae of this section.

In the full theory, where the $\eta'$ is massive, one gets
\begin{equation}
\chi_{tL}(0)=0\, ,
\label{AWTI3}
\end{equation}
which is the (lattice) regularized version of eq.~(\ref{SUM}).

The key observation at this point is that, at vanishing quark mass, in the 
limit $u\rightarrow 0$, only the $\eta'$ pole contributes to $\chi_{tL}(p)$, 
as explained below eq.~(\ref{ETAQ}), giving (see eqs.~(\ref{CHIYM0}) 
and~(\ref{RETAF}) of the Appendix)
\begin{equation}
\lim_{p\rightarrow 0}\lim_{u\rightarrow 0} \chi_{tL}(p)=
\frac{F_\pi^2}{2N_f} m_{\eta'}^2\Big{|}_{u=0}\, .
\label{AWTI4}
\end{equation}
Recalling that $F_\pi={\rm{O}}(\sqrt{N_c})$~\cite{NINF,VENTOP} to 
leading order in $u$, we see that consistently the combination in the r.h.s. 
of this equation has a finite limit as $u\rightarrow 0$.

On the other hand, using the WTI~(\ref{AWTIL}) with $\hat{O}=\hat{Q}$, 
we can put eq.~(\ref{AWTI2}) in the form
\begin{eqnarray}
&\!\!\!\!\!\!\!&\chi_{tL}(p)=\label{AWTI5}\\&\!\!\!\!\!\!\!&=(1-Z)\int d^4x\, 
{\rm{e}}^{-ipx} \<\frac{1}{2}{\rm{Tr}}[\gamma_5 D(x,x)]
\Big{(} \frac{1}{2}{\rm{Tr}}[\gamma_5 D(0,0)]-
\frac{Z}{2N_f}\nabla_\mu A^0_\mu (0)\Big{)}\>\, ,\nn
\end{eqnarray}
since $\delta^x_A\hat{Q}=0$. Taking, as before, the limit $p\rightarrow 0$ 
after $u\rightarrow 0$ in eq.~(\ref{AWTI5}), we get
\begin{eqnarray}
&\!\!\!\!\!\!\!\!\!\!\!\!&\frac{F_\pi^2}{2N_f} m_{\eta'}^2\Big{|}_{u=0}=
\lim_{p\rightarrow 0}\lim_{u\rightarrow 0}\Big{[}
(1-Z)\int d^4x\, {\rm{e}}^{-ipx}\<\frac{1}{2}{\rm{Tr}}[\gamma_5 D(x,x)]
\frac{1}{2}{\rm{Tr}}[\gamma_5 D(0,0)]\>+ 
\nn\\ &\!\!\!\!\!\!\!\!\!\!\!\!& -
\frac{Z}{1-Z}\frac{F_\pi^2}{2N_f}m_{\eta'}^2\Big{]}\, ,
\label{AWTI6}
\end{eqnarray}
where eq.~(\ref{AWTI4}) has been used in the l.h.s. and the last term comes 
from the contribution of the $\eta'$ pole to the second term in the r.h.s. 
of eq.~(\ref{AWTI5}). Recalling that $Z$ vanishes when $u\rightarrow 0$, 
we obtain
\beq
\frac{F_\pi^2}{2N_f} m_{\eta'}^2\Big{|}_{u=0}=
\lim_{p\rightarrow 0}\lim_{u\rightarrow 0}
\int d^4x \,{\rm{e}}^{-ipx}\<\frac{1}{2}{\rm{Tr}}[\gamma_5 D(x,x)]
\frac{1}{2} {\rm{Tr}}[\gamma_5 D(0,0)]\>\, .
\label{ETAINTER}\eeq
Under the smooth-quenching hypothesis the limits $u\rightarrow 0$ and 
$p\rightarrow 0$ in this expression can be readily performed in the order 
indicated, as the first one simply amounts to 
setting the fermion determinant equal to unity. One gets in this way
\beq
\frac{F_\pi^2}{2N_f} m_{\eta'}^2\Big{|}_{u=0}=
\int d^4x \,\<\frac{1}{2}{\rm{Tr}}[\gamma_5 D(x,x)]
\frac{1}{2} {\rm{Tr}}[\gamma_5 D(0,0)]\>\Big{|}_{\rm{YM}}\, .
\label{ETAMASSOVER}\eeq
The restriction to the pure YM theory, indicated in eq.~(\ref{ETAMASSOVER}), 
is an obvious consequence of the fact that, for a Green function with the 
insertion of only gluonic operators, neglecting the fermion determinant 
is tantamount to limiting the functional integral to the pure gauge 
sector of the theory. 

A relevant question at this point is to ask for which value of $N_c$ 
eq.~(\ref{ETAMASSOVER}) is supposed to be valid. The answer depends on the 
detailed behaviour of QCD with $N_f$. Various possibilities are logically 
envisageable. 

1) In the most favorable situation, in which the limit $u\rightarrow 0$ of 
$\chi_{tL}(p)$ exists and is equal to the value it takes at $N_f=0$ 
(fermion determinant equal to 1), the formula~(\ref{ETAMASSOVER}) is valid 
for any value of $N_c$ and for each $N_c$ it yields the mass of the 
$\eta'$ meson to leading order in $u$ in the world with the corresponding 
number of colours, without any O($1/N_c$) corrections~\cite{VEN}. 

2) If quenching can be attained only in the limit in which the number of 
colours goes to infinity, then eq.~(\ref{ETAMASSOVER}) will yield a formula 
for the $m_{\eta'}^2$ valid up to O($1/N_c$) corrections~\cite{WIT}.

3) Finally it may happen that taking $u\rightarrow 0$ does not correspond to
quenching. In this case one cannot pass from the fairly complicated 
eq.~(\ref{ETAINTER}) to the more useful formula~(\ref{ETAMASSOVER}). 

Equation~(\ref{ETAMASSOVER}) has precisely the same structure 
as the naive formula~(\ref{ETAMASS}), when the expression~(\ref{A}) of the 
YM topological susceptibility is inserted in it. Remembering the  
meaning of $\frac{1}{2}\int d^4x {\rm{Tr}}[\gamma_5 D(x,x)]$ 
(eq.~(\ref{INDEX})), eq.~(\ref{ETAMASSOVER}) can be rewritten in the 
suggestive form
\beq
\frac{F_\pi^2}{2N_f} m_{\eta'}^2\Big{|}_{u=0}=
\lim_{V\rightarrow\infty}\frac{\<(n_R-n_L)^2\>}{V}\, ,
\label{ETAMASSIND}\eeq
where $\<(n_R-n_L)^2\>$ is the expectation value of the square of the 
index of the GW fermion operator, $D$, and $V$ is the physical volume 
of the lattice.

It may be interesting to see what are the implications of the hypothetical 
situation in which the $\eta'$ mass would not vanish (linearly) with $u$. 
In this case the two limits $u\rightarrow 0$ and $p\rightarrow 0$ of 
$\chi_{tL}$ would commute, leading to the chain of relations
\begin{equation}
0=\lim_{u\rightarrow 0}\lim_{p\rightarrow 0} \chi_{tL}(p)=
\lim_{p\rightarrow 0}\lim_{u\rightarrow 0} \chi_{tL}(p)=
\lim_{V\rightarrow\infty}\frac{\<(n_R-n_L)^2\>}{V}\geq 0\, ,
\label{PRO}
\end{equation}
which imply either that the non-negative quantity $(n_R-n_L)^2$ vanishes 
identically for any gauge configuration or, at least, that the quantity 
$\<(n_R-n_L)^2\>$ grows less than linearly with the number of lattice points.

\section{Numerical results}
\label{sec:NUM}

In the form~(\ref{ETAMASSIND}) the $\eta'$ mass formula can be directly
compared with overlap fermion simulation data, because the problem is 
reduced to counting the number of zero modes of the overlap-Dirac operator and
measuring their chirality. There exist data both in 2 (abelian Schwinger 
model~\cite{GHR}) and in 4 ($SU(3)$ gauge theory~\cite{EHN}) dimensions.

\subsection{2 dimensions: abelian Schwinger model}
\label{subsec:TWOD}

The validity of the WV formula can be explicitly checked in the case of the
two-dimensional abelian Schwinger model~\cite{SCH}, since the model 
is exactly soluble~\cite{SCH,SEI}. Numerical checks are also possible, as 
there exist very recent simulations that make use of the overlap fermion
formalism~\cite{GHR}.

For the quantities of interest, one finds the formulae~\cite{SS}
\beq
\hat Q(x)=\frac{e}{4\pi}\epsilon_{\mu\nu}F_{\mu \nu}
\label{Q2}\eeq
\beq
\chi_{t}(p)=\frac{e^2}{4\pi^2}\Big{(}1-
\frac{e^2/\pi}{p^2+e^2/\pi}\Big{)}\, .
\label{G2}\eeq
These formulae are for $N_f=1$. The generalization to the case with
$N_f$ species of fermions can be found at the end of the Appendix. 

By inspection one immediately recognizes that $\chi_{t}(p=0)=0$ and that 
there exists one single particle in the spectrum, a boson (the analogue of 
the $\eta'$ in QCD) with mass
\beq
m_{\eta'}^2=\frac{e^2}{\pi}\, .
\label{M2}\eeq
The residue of the $\eta'$ pole is given by~\cite{SS}
\begin{equation}
R^2_{\eta'}=\Big{|}\<0|\hat Q|{\eta'}\>\Big{|}^2 =
\Big{(}\frac{e^2}{\pi\sqrt{4\pi}}\Big{)}^2\, .
\label{RES2}
\end{equation}
As the negative term in eq.~(\ref{G2}) is fully contributed by the fermion 
determinant, the $\eta'$ mass formula in 2 dimensions reads
\begin{equation}
\frac{R^2_{\eta'}}{m^2_{\eta'}} =  
\chi_{t}(0)\Big{|}_{\rm{quenched}}=\frac{e^2}{4\pi^2}\, .
\label{MASSD2}
\end{equation}
Comparing eq.~(\ref{MASSD2}) with~(\ref{RES2}), one immediately checks the 
consistency of~(\ref{MASSD2}) with~(\ref{M2}). Incidentally, looking back
at eq.~(\ref{ETAMASSOVER}), we conclude that $F_{\eta'}=1/\sqrt{2\pi}$.

Numerically one finds $\<(n_R-n_L)^2\> = 2.7\pm 0.4$, at $1/e^2=6$ 
in a lattice with $N=24^2$ points~\cite{GHR}, implying
\begin{equation}
\frac{\<(n_R-n_L)^2\>}{N} =
\frac{2.7\pm 0.4}{24^2}= (4.7\pm 0.7)\times 10^{-3}\, .
\label{IND2}
\end{equation}
Although this number compares fairly well with the expected result 
(eq.~(\ref{MASSD2}))
\begin{equation}
\frac{e^2}{4\pi^2}=\frac{1/6}{4\pi^2 }= 4.2\times 10^{-3}\, ,
\label{EXAC2}
\end{equation}
further simulations at different $\beta$'s and larger volumes are needed to 
have properly under control all sources of systematic errors.

\subsection{4 dimensions: $SU(3)$ gauge theory}
\label{subsec:FOURD}

Using the spectral flow method~\cite{NN}, the expectation value of
$(n_R-n_L)^2$ in an $SU(3)$ gauge theory has been computed for three lattice
volumes ($N=8^3\times 12$, $8^3\times 16$, $16^3\times 8$) at $\beta=5.85$
in the first paper of ref.~\cite{EHN}. The data show a fairly good scaling
with $N$. If the value of the string tension $\sigma=(440 \pm 38\, 
{\rm{MeV}})^2$~\cite{TEP} is used together with 
$a\sqrt{\sigma}=0.2874(7)$~\cite{EHK} 
to fix the lattice spacing, one finds ($V=Na^4$)
\beq
A=\frac{\<(n_R-n_L)^2\>}{V}=\Big{(}198\pm 20\, {\rm{MeV}}\Big{)}^4\, ,
\qquad \beta=5.85\, .
\label{ETAMASSNUM}\eeq
The same authors~\cite{EHN} have also produced data at larger values
of $\beta$ and volumes. At $\beta=6.0$ and $N=16^3\times 32$, using 
$a\sqrt{\sigma} = 0.2189(9)$~\cite{EHK}, they get
\beq
A=\frac{\<(n_R-n_L)^2\>}{V}=\Big{(}194\pm 20\, \mbox{MeV}\Big{)}^4\, ,
\qquad \beta=6.0\, .
\label{ETAMASSNUM1}\eeq
Errors in eqs.~(\ref{ETAMASSNUM}) and~(\ref{ETAMASSNUM1}) have been computed
by combining in quadrature the uncertainties on $\<(n_R-n_L)^2\>$, $\sigma$ and
$a\sqrt{\sigma}$. It is remarkable that the central value of $A$
is quite stable with $\beta$ and very near to the number
required to match the actual value of the $\eta'$ meson mass~\cite{VEN}.
Clearly more work is needed (larger $\beta$ values and volumes) before
these encouraging indications are fully confirmed and the
magnitude of the error is properly assessed.

\section{Conclusions}
\label{sec:CONCL}

In this paper we have given a derivation of the WV formula for the $\eta'$ 
mass in lattice QCD, by exploiting the flavour-singlet 
axial WTIs of the theory. If fermions obeying the GW relation~(\ref{GWREL}) 
are used, there exists a natural definition of topological density 
(eq.~(\ref{CHARDEN})), for which the naive form of the WV formula holds 
without the need of introducing any subtraction. 

The validity of a formula of the WV type for the $\eta'$ mass,
i.e.~of a formula that is telling us that the O($\Lambda_{_{\rm{QCD}}}$)
contribution to $m_{\eta'}$ originates from the breaking of the $U_A(1)$ 
symmetry due to the gluon anomaly, is a key test of our understanding of 
strong interaction dynamics, as described by QCD. It is reassuring that 
such a formula holds even beyond the formal continuum-like derivations
given in refs.~\cite{WIT,VEN}.

In the literature essentially two kinds of approaches have been proposed 
to deal with the problem of computing of the topological susceptibility 
on the lattice, which have led to numerical values for $A$ as good as those 
obtained here. The first one is based on a direct field theoretical 
definition of $A$~\cite{DIG} that takes into account the need for the 
renormalization of $Q_L$ and the subtraction of the operators $F^2$ and 
$1\!\!\!\! 1$ in the short distance expansion of $Q_LQ_L$. The 
second one makes use of the notion of ``cooling''~\cite{DFGPS} to carry out 
the necessary operations of renormalization and subtraction. Both methods are 
reliable to the extent that they are able to capture the topological 
properties of the gauge field configurations that determine the number of 
zero modes of the Dirac operator. Simulations based on the geometrical 
definition of ref.~\cite{LU1} have not yet led to comparably good 
results~\cite{SCHI}.

We conclude by noting that a completely analogous set of arguments,
as those developed in this paper, could be carried out for the pseudo-scalar 
meson (sort of $\eta/\pi$ particle) belonging to the lightest
supermultiplet of the ${\cal{N}}=1$ Super-YM gauge theory, the only 
difference being that in this case one would be dealing with Weyl fermions 
in the adjoint representation of the gauge group. In order to 
assess the numerical importance of gluino loops for the restoration 
of supersymmetry, it would be interesting to compare the values of the 
masses of the other two partners of the lowest lying supermultiplet, as 
they are obtained in quenched simulations~\cite{MONT1}, with the number 
one would derive for the mass of the $\eta/\pi$ particle following the 
present analysis.
\vskip .3cm

{\bf Acknowledgements} - We wish to thank M. L\"uscher for his interest in
this work and for illuminating discussions. Discussions with D. Anselmi, 
A. Di Giacomo, A. Hasenfratz, P. Hasenfratz, P. Hern\'{a}ndez, C. Hoelbling, 
F. Niedermayer, C. Rebbi and A. Vladikas and correspondence with R. Crewther, 
G.M. Shore and G. Travaglini are also gratefully acknowledged. L.G. was partly 
supported by DOE grant DE-FG02-91ER40676. G.V. wishes to acknowledge the 
support of a ``Chaire Internationale Blaise Pascal", administered by the 
``Fondation de L'Ecole Normale Sup\'erieure'', during most of this work.

\vspace{-.4cm}
\section*{Appendix}

We present here a simple (and to our knowledge novel) derivation of 
a number of relations between certain hadronic observables and various 
quantities related to (the continuum limit of) the ``topological 
susceptibility correlator'' $\chi_t(p)$ (see eq.~(\ref{AWTI2})).
\vspace{-.4cm}
\subsection*{The dispersion relation for $\chi_t(p)$}

As $\chi_t(p)$ is a quantity of mass dimension 4, it will satisfy a 
three-times-subtracted dispersion relation~\footnote{The same results
as those obtained below follow, also if one starts with a dispersive
integral having a smaller number of subtractions.} that we write in the form
\beq
\chi_t(p)=\chi_t(0)+\chi_t'(0)p^2+\frac{1}{2}\chi_t''(0) (p^2)^2+ 
(p^2)^3 I\, ,
\label{DISPREL} \eeq
where for a properly defined correlator we have (see eq.~(\ref{AWTI3}))
\beq
\chi_t(0)=0 \, . 
\label{SUMRULE} \eeq
In eq.~(\ref{DISPREL}) $I$ represents the dispersive integral involving 
${\rm{Im}}\chi_t(p)$. It is useful to express $I$ by separating out 
the $\eta'$ contribution from the rest. We thus put
\beq
I\equiv \frac{1}{(m_{\eta'}^2)^3}
\frac{R^2_{\eta'}}{p^2+m_{\eta'}^2} +\tilde{I}\, ,
\label{IETA} \eeq
where $-R^2_{\eta'}$ is the full physical residue of the $\eta'$ pole, 
which, as explicitly indicated, is negative-definite in Euclidean metric. 
This is a consequence of the fact that, by analytical continuation 
from the Minkowski region~\cite{SS}, the whole dispersive integral 
is negative-definite. This circumstance solves the famous sign 
problem discussed in~\cite{WIT}.

The key observation at this point is that for the $\eta'$ contribution the
limits $p\rightarrow 0$ and $u\rightarrow 0$ do not commute, since (if)
$m_{\eta'}^2={\rm{O}}(u)$. We then proceed by computing the two sides of 
eq.~(\ref{DISPREL}) by taking $u\rightarrow 0$ before $p\rightarrow 0$.

In doing so, in the l.h.s. of eq.~(\ref{DISPREL}) we simply get the quantity
$\chi_t(p)$ in the quenched theory, i.e.
\beq
\chi_t(p)|_{\rm{quenched}}=\lim_{u\rightarrow 0}\chi_t(p)\, .
\label{CHIQ} \eeq
This follows from the assumption that, since $p\neq 0$, taking the limit 
$u\rightarrow 0$ (before $p\rightarrow 0$) is equivalent to neglecting the 
contribution of the fermion determinant (smooth-quenching hypothesis). In the 
r.h.s. we expand at fixed $p^2$ the $\eta'$ contribution in powers of 
$m_{\eta'}^2/p^2={\rm{O}}(u/p^2)$ and then we match the resulting $p^2$ 
power behaviour with the Taylor expansion of $\chi_t(p)|_{\rm{quenched}}$ 
around $p^2=0$. We get in this way the relations
\beq
\chi_t''(0)|_{\rm{quenched}}=\lim_{u\rightarrow 0}
\left[\chi_t''(0)+2\frac{R^2_{\eta'}}{m_{\eta'}^6}\right]\, ,
\label{CHIYM2} \eeq
\beq
\chi_t'(0)|_{\rm{quenched}}=\lim_{u\rightarrow 0}
\left[\chi_t'(0)-\frac{R^2_{\eta'}}{m_{\eta'}^4}\right]\, ,
\label{CHIYM1} \eeq
\beq
\chi_t(0)|_{\rm{quenched}}=\lim_{u\rightarrow 0}
\left[\chi_t(0)+\frac{R^2_{\eta'}}{m_{\eta'}^2}\right]
\, .
\label{CHIYM0} \eeq
Since $\tilde I$ is a smooth function of $u$ and $p$, which vanishes as 
$p\rightarrow 0$ with its first two derivatives, it does not give any 
contribution to the above equations.

The smooth-quenching hypothesis implies that the combinations appearing 
in the r.h.s. of eqs.~(\ref{CHIYM2}) to~(\ref{CHIYM0}) have a finite limit 
when $u\rightarrow 0$. As a consequence, they can be rewritten in the 
more expressive form
\beq
\frac{R^2_{\eta'}}{m_{\eta'}^2}=\chi_t(0)|_{\rm{quenched}}-\chi_t(0)
+{\rm{O}}(u)\, ,
\label{CHIT0} \eeq
\beq
-\frac{R^2_{\eta'}}{m_{\eta'}^4}=\chi_t'(0)|_{\rm{quenched}}- \chi_t'(0)
+{\rm{O}}(u)\, ,
\label{CHIT1} \eeq
\beq
2\frac{R^2_{\eta'}}{m_{\eta'}^6}= \chi_t''(0)|_{\rm{quenched}}- \chi_t''(0)
+{\rm{O}}(u)\, .
\label{CHIT2} \eeq
In this way it clearly appears that, to leading order in $u$, the physical 
quantities in the l.h.s. of these equations are not affected by the presence 
of any possible counter-term, introduced to make the topological 
susceptibility correlator finite at $p\neq 0$ (see the discussion after 
eq.~(\ref{GQQNC})). This immediately follows from the observation that 
the $u$-independent part of such counter-term, even if non zero, will cancel 
between quenched and unquenched quantities in the r.h.s. of 
eqs.~(\ref{CHIT0}), (\ref{CHIT1}) and~(\ref{CHIT2}).

\subsection*{The $\eta'$ mass formula}

Owing to eq.~(\ref{SUMRULE}), eq.~(\ref{CHIT0}), taken in the limit 
$u\rightarrow 0$, is nothing but the WV relation. To see this  
we recall that the residue of the $\eta'$ pole is given by the formula
\begin{equation}
R^2_{\eta'}=\Big{|}\<0|\hat Q|{\eta'}\>\Big{|}^2\, .
\label{RES}
\end{equation}
To get an explicit expression of $R^2_{\eta'}$, it is sufficient to write 
the matrix element of $\hat Q$ appearing in eq.~(\ref{RES}) in terms 
of the corresponding matrix element of the divergence of the flavour-singlet 
axial current, using the anomalous PCAC relation. As we are in the 
chiral limit, we get
\beq
2N_f \<0|\hat Q|{\eta'}\>= \<0|\nabla^\mu \hat A_\mu^0|{\eta'}\>\, .
\label{PCAC}\eeq
We can express the r.h.s. of eq.~(\ref{PCAC}) in terms of $F_\pi$, since, 
for the sake of deriving the WV formula, we are actually only 
interested in computing $R^2_{\eta'}$ in the limit $u\rightarrow 0$, where 
the pseudo-scalar meson decay constants, $F_\pi$ and $F_{{\eta}'}$, coincide.
Combining the equation~\footnote{The numerical factor $\sqrt{2N_f}$ in  
equation~(\ref{FPI}) needs some explanation. The factor $\sqrt{2}$ comes 
from the fact that $F_\pi$ is defined through the equation 
$\<0|\nabla^\mu A_\mu^{\pi^0}|\pi^0\>=F_\pi m^2_{\pi}$, 
where $\pi^0=2^{-1/2}(\bar u u -\bar d d)$, $A_\mu^{\pi^0} =
\bar q \gamma_\mu\gamma_5 \lambda^{\pi^0} q$ and $\lambda^{\pi^0}=
\lambda_3/2$, with $\lambda_3$ the usual Gell-Mann matrix. The factor 
$\sqrt{N_f}$ is due to the fact that the flavour-singlet axial current, 
$A_\mu^{0}$, is a sum over all light flavours with weight 1, while the 
wave function of the flavour-singlet meson has a $1/\sqrt{N_f}$ factor 
when expressed in terms of $\bar q q$ states. With these definitions 
$F_\pi$ is of O(1) in $N_f$ to leading order in $N_c$.}
\beq 
\<0|\nabla^\mu \hat A_\mu^0|{\eta'}\>=\sqrt{2N_f}F_{\eta'}
m^2_{\eta'}\, ,
\label{FPI} \eeq
which defines $F_{\eta'}$, with~(\ref{PCAC}), we obtain
\beq
\frac{R^{2}_{\eta'}}{m^2_{\eta'}}\Big{|}_{u=0}
=\frac{F_{\eta'}^2 m^2_{\eta'}}{2 N_f}\Big{|}_{u=0}
=\frac{F_\pi^2 m^2_{\eta'}}{2 N_f}\Big{|}_{u=0}\, . 
\label{RETAF} \eeq
Recalling that $F_\pi={\rm{O}}(\sqrt{N_c})$~\cite{NINF,VENTOP} to 
leading order in $u$, we see that we consistently have 
$R^2_{\eta'}/m_{\eta'}^2|_{u=0}={\rm{O}}(1)$.
If we now put together eqs.~(\ref{CHIT0}) and~(\ref{RETAF}), we finally get
\beq
m^2_{\eta'}=\frac{2 N_f}{F_\pi^2}\chi_t(0)|_{\rm{quenched}}
+{\rm{O}}(u^2)\, .
\label{ETAPMAS} \eeq

It is amusing to notice that eliminating $R^2_{\eta'}/m_{\eta'}^2$ 
between eqs.~(\ref{CHIT1}) and~(\ref{CHIT2}) provides a new 
formula for the $\eta'$ mass, namely
\beq
m_{\eta'}^2=-2\,\frac{\chi_t'(0)|_{\rm{quenched}}-\chi_t'(0)}
{\chi_t''(0)|_{\rm{quenched}}-\chi_t''(0)} +{\rm{O}}(u^3)\, ,
\label{CHIETA} \eeq
which, as indicated above, is more accurate than the WV 
formula~(\ref{ETAPMAS}). It should be stressed that from eq.~(\ref{CHIETA}) 
$m_{\eta'}^2$ is expressed, up to O($u^3$), in terms of differences between 
``quenched'' and ``full'' quantities that are not affected by the presence 
of possible counter-terms in the definition of the topological susceptibility 
correlator at $p\neq 0$.

\subsection*{A formula for $\chi'_t(0)$}

An interesting formula for $\chi'_t(0)$ can be derived from 
eq.~(\ref{CHIT1}). In fact, recalling eq.~(\ref{CHIT2}), the former can be 
identically rewritten in the more expressive fashion
\beq
\chi_t'(0)=\frac{F_\pi^2}{2 N_f}+
\left[\frac{R^2_{\eta'}}{m_{\eta'}^4} - \frac{F_\pi^2}{2 N_f} 
+\chi_t'(0)|_{\rm{quenched}}\right] +{\rm{O}}(u)\, ,
\label{CHIEJ} \eeq
where the first term is O($1/u$) and the term in square brackets is of O(1), 
as follows from eq.~(\ref{RETAF}). An expansion of this kind for $\chi_t'(0)$ 
can be useful in connection with the discussion of the proton spin crisis 
problem, given in ref.~\cite{PSCP}, where it was noticed that, in the limit 
in which $\chi_t'(0)$ is identified with ${F_\pi^2}/{2 N_f}$, one 
recovers the Ellis--Jaffe sum rule~\cite{EJ}. It would be interesting 
to see whether the correction in square brackets goes in the right direction 
for solving the proton spin puzzle.

For the study of the proton spin problem one is directly interested in 
$\chi'_t(0)$ itself and not in ``quenched''--``full'' differences, like 
the combinations~(\ref{CHIT0}), (\ref{CHIT1}) and~(\ref{CHIT2}) 
appearing in the formulae~(\ref{ETAPMAS}) and~(\ref{CHIETA}) for the $\eta'$ 
mass. Thus the question arises whether the quantity $\chi'_t(0)$ can be 
defined in a way amenable to unambiguous numerical simulations. This is the 
case only if terms proportional to $p^2$ are absent in the polynomial 
expansion of the counter-term, CT($p$). However, in a (lattice) regularized 
theory the presence of a quadratically divergent term of the type 
$p^2 1\,\!\!\!1/a^2$ cannot be excluded and will thus have to be 
compensated for. This means that there is no obvious way to define 
$\chi_t'(0)$ in a way suitable for numerical simulations, at least from 
these simple considerations only.

\subsection*{The two-dimensional case}

We conclude by noticing that in the two-dimensional abelian ($N_c=1$) 
Schwinger model with $N_f$ species of fermions, one can directly verify 
the validity of eqs.~(\ref{CHIT0}) to~(\ref{CHIT2}), which, in fact,
hold exactly without O($u=N_f$) corrections. It is sufficient to observe
that the generalization of eq.~(\ref{G2}) to arbitrary $N_f$ reads
\beq
\chi_{t}(p)=\frac{e^2}{4\pi^2}\left(1-
\frac{N_f e^2/\pi}{p^2+N_f e^2/\pi}\right)\, ,
\label{GN2}\eeq
while the ratio~(\ref{MASSD2}) is independent of $N_f$.

\end{document}